\documentclass[12pt, eqsecnum]{revtex4}
\begin{document}
\def\gtap{\mathrel{ \rlap{\raise 0.511ex \hbox{$>$}}{\lower 0.511ex
   \hbox{$\sim$}}}} \def\ltap{\mathrel{ \rlap{\raise 0.511ex
   \hbox{$<$}}{\lower 0.511ex \hbox{$\sim$}}}}
\newcommand{\beq}{\begin{equation}}
\newcommand{\dd}{\partial}
\newcommand{\eeq}{\end{equation}}
\newcommand{\bea}{\begin{eqnarray}}
\newcommand{\eea}{\end{eqnarray}}
\newcommand{\lsim}{\stackrel{<}{\scriptstyle \sim}}
\newcommand{\gsim}{\stackrel{>}{\scriptstyle \sim}}
\newcommand{\La}{{\cal L} }
\newcommand{\half}{\frac{1}{2} }
\newcommand{\eq}[1]{eq.(\ref{#1})}
\newcommand{\dpar}[2]{\frac{\partial #1}{\partial #2}}
\newcommand{\vpar}[2]{\frac{\delta #1}{\delta #2}}
\newcommand{\ddpar}[2]{\frac{\partial^2 #1}{\partial #2^2}}
\newcommand{\vvpar}[2]{\frac{\delta^2 #1}{\delta #2^2}}
\newcommand{\eV}{\mbox{$ \ \mathrm{eV}$}}
\newcommand{\KeV}{\mbox{$ \ \mathrm{KeV}$}}
\newcommand{\MeV}{\mbox{$ \ \mathrm{MeV}$}}
\newcommand{\probm}{\mbox{$ \ \langle P_m \rangle$}}
\newcommand{\bz}{\bar{z}}
\newcommand{\sg}{\sigma}
\newcommand{\tw}{\tilde{\omega}}
\newcommand{\rla}{\longrightarrow}


\title{ New QES Hermitian as well as non-Hermitian PT invariant Potentials }

\author{ Avinash Khare and Bhabani Prasad Mandal }
\affiliation{ Institute of Physics, Sachivalaya Marg, Bhubaneswar-751005,
 India} 
\affiliation{  {Department of Physics, Banaras Hindu University, Varanasi-221005, India}}


\begin{abstract}
We start with quasi-exactly solvable (QES) Hermitian (and hence real) as well
as complex PT-invariant, double sinh-Gordon potential 
and show that even after adding perturbation terms, the resulting 
potentials, in both cases, are still QES potentials. 
Further, by using anti-isospectral transformations, we obtain 
Hermitian as well as PT-invariant complex QES periodic potentials.
We study in detail the various properties of the corresponding 
Bender-Dunne polynomials. 
\end{abstract}

\maketitle
\section{Introduction}
In recent years, various features of the complex PT-symmetric Hamiltonians
have been explored in the literature,see for example Refs.
 \cite{bed,ali,oth,bpm}. In particular, it has
been shown that so long as the PT-symmetry is not spontaneously broken, then 
the energy eigenvalues of the Schr\"odinger equation are real. Further,
several PT-symmetric complex QES\cite{ush,bt} potentials have been
 discovered \cite{km1,km,qes1}. The purpose
of this paper is to point out that even after suitably perturbing either 
a Hermitian or a complex
PT-invariant QES potential, one can still obtain another Hermitian or complex, 
PT-invariant, QES potential.

\section{ PT symmetric Non-Hermitian QES system}

\subsection {The Model}

We start from the well known non-Hermitian but PT-invariant
potential $-[a\cosh (2x)-iM)^2\,,\, (\hbar =1 = 2m $) \cite{km1}
and show that even the perturbed complex PT-invariant Hamiltonian
\beq
H= p^2 - [a\cosh (2x) -iM]^2 +\frac{l(l+1)}{\sinh^2(x)}
-\frac{l(l+1)}{\cosh^2(x)},\label{1}
\eeq 
 where $a,l, M $ are real and  $-1<l<0 $ is a QES potential so long as
$M-2l-1$ or $M+2l+1$ is a positive even integer. Note, the restriction
$-1<l<0$ has been imposed so that
the potential is not too singular at $x=0$ and there is a communication from
the left side to the right side.  

Let us first show that the above non-Hermitian Hamiltonian (\ref{1}) 
is a PT-symmetric one. We note that in this case the parity transformation
is defined as $ x\longrightarrow i\frac{\pi}{2} -x $, which is reflection
of coordinate about the point $x= i\frac{\pi}{4} $. Under the time reversal 
transformation $t\longrightarrow -t$ and further, one replaces 
$ i \longrightarrow -i $. One can check easily that, under PT 
$\cosh(2x)\rla -\cosh(2x)$ ,$\cosh^2(x)\rla -\sinh^2(x) $ and
$ \sinh^2(x)\rla -\cosh^2(x) $.

 We substitute
 \beq
\psi(x) = e^{i\frac{a}{2}\cosh (2x)}\phi(x) \,\label{2}
\eeq 
in the Schro$\ddot{o}$dinger equation  $H\psi= E\psi$ with the H given
by Eq. (\ref{1}) and obtain 
\bea\label{2a}
&&\phi^{\prime\prime}(x) +2ia\sinh (2x)\phi^\prime (x) + 
[(E-M^2+a^2) \nonumber \\ && 
-2i(M-1)a\cosh (2x) 
-\frac{l(l+1)}{\sinh^2(x)}
 +\frac{l(l+1)}{\cosh^2(x)} ]\phi(x) =0 \nonumber\,  
\eea 

On further substituting
 \beq\label{3}
\phi = [\cosh(x)]^{\alpha} [\sinh(x)]^{\beta} \eta (x)\,,
\eeq 
we find that the system admits non-singular QES solutions provided
\beq
\alpha(\alpha+1) = l(l+1) \mbox{and }\ \beta(\beta+1) = l(l+1)
\label{n1}\eeq
is satisfied.
 The condition in Eq. (\ref{n1}) implies either
of the following four conditions  i.e. 

(i) $\alpha=\beta=l+1 $ (ii) $\alpha=\beta=-l $ (iii) $ \alpha=-l, \beta=l+1$ 
(iv) $ \alpha=l+1, \beta=-l$.

We discuss all the cases separately,

{\bf Case (i)}: $\alpha=\beta=l+1$,\ \ 

In this case, $\eta (x)$ as given by Eq. (\ref{3}) can be shown to satisfy
\bea\label{4}
&&\eta^{\prime\prime}(x) +2[(l+1)\coth(x)+(l+1) \tanh(x)+ia\sinh (2x)]
\eta^\prime (x) \nonumber \\
&&+ \left[\epsilon-iza\cosh^2 (x) \right ] \eta(x) =0 \,,  
\eea 
where
\bea\label{5}
\epsilon &= & E-M^2+a^2+4(l+1)^2-2ia(2l-M+3)\,\nonumber \\
  z &= & 4(M-2l-3)\,.
\eea 
 This system has $p$ QES solutions in case $M-2l-1=2p$  with $p=1,2,3,\cdots $.
 These solutions are of the form 
\beq\label{6}
\eta (x) = \sum_{n=0}^{\infty} a_n (i)^{n}[\cosh2(x)]^{n}\,,
\eeq 
 Two of the low lying solutions are 
\beq\label{7}
M=2l+3\,,~~\eta = constant\,,~~ E=-a^2+4l+5\,,
\eeq 
\bea\label{8}
&&M=2l+5\,,~~ \eta=A\cosh(2x)+iB\,,~~\frac{A}{B}=\frac{E+a^2-12l-21}{4a}\,,
 \nonumber \\
&&E=8l+15-a^2\pm \sqrt{(2l+3)^2-4a^2}\,.
\eea 
Note that since $\alpha =\beta $ the PT symmetry is unbroken and hence, as
expected, the energy eigenvalues are real.

{\bf Case (ii)}:\ $\alpha =\beta =-l$. 

The solutions in this case can be obtained from the solutions
of case (i) by everywhere changing $l \longrightarrow -l-1$. Thus 
in this case, one
has $p$ QES solutions in case $M+2l+1=2p$ with $p=1,2,3,\cdots $.
Note that since $\alpha =\beta $, even in this case, the PT symmetry is 
unbroken and the energy eigenvalues are real.

{\bf Case (iii)}:\ $\alpha=-l\,,~\beta=l+1$.

In this case we have  $p$ QES solutions in case  $M=2p,\, p=1,2, \cdots$ .
Two of the low lying QES solutions are
\beq\label{9} 
M=2\,,~~  E=3-a^2-2ai(1+2l)\,, 
\eeq
\beq\label{10}
M=4\,,~~E=11-a^2-2ai(1+2l)\pm \sqrt{(1-a^2)-ia(1+2l)}\,.
\eeq 
Note that since in this case  $\alpha \ne \beta $, hence the PT symmetry is 
broken spontaneously and eigenvalues are no more real.
In the special case of $\alpha = \beta = \frac{1}{2}$,
the PT symmetry is restored and $E$ becomes real.

{\bf Case (iv)}: In this case, the solutions can simply be obtained from the 
solutions of the case (iii) by changing $l\longrightarrow -l-1$.

Ordinarily, the boundary conditions that give the quantized energy
levels are $\psi(x)\rightarrow 0$ as $|x|\rightarrow \infty $
on the real axis. However, in the present case, we have to continue the
eigenvalue problem into the complex $-x$ plane\cite{bt}. On putting,
$x=u+iv$ where $u$ and $v$ are real, it is easy to see that  for $u>0$
 the boundary condition is satisfied  so long
 as $-\pi<v<-\frac{\pi}{2}$ (mod $\pi $) while for $u<0$ it is satisfied if $-\frac{\pi}{2}<v<0$
(mod $\pi $)

It is worth mentioning that, the non-Hermitian PT symmetric
Hamiltonian in Eq. (\ref{1}) after a suitable change of variable
 can be expressed  
in terms of the $SL(2,R)$ generators ( at most quadratic).
To show this we substitute $t=\cosh (2x)$ in the Eq. (\ref{4}) to obtain
$H_g\eta =E\eta $ where,
\begin{eqnarray}
H_g= -4(t^2-1)\frac{d^2}{dt^2}-[(8l+12)t &+& 4ai(t^2-1)]\frac{d}{dt}-[-M^2+a^2
\nonumber\\ &+&4(l+1)^2-2ai(2l-M+3) 
-\frac{iz}{2}(t+1)]
\end{eqnarray}
This gauged Hamiltonian then can be expressed in terms of the generators
of the $SL(2,R)$ by
\begin{eqnarray}
H_g=-4[(J_0^2-J_-^2)+ia(J_+-J_-) &+&(n+2l+2)J_0]- [-M^2+a^2\nonumber \\ 
&+& 4(l+1)^2+4n(l+1)+n^2]
\end{eqnarray}
while the generators are given by
$J_-=\frac{d}{dt},\ J_0= t\frac{d}{dt}-\frac{n}{2}$ and 
$ J_+= t^2\frac{d}{dt}-nt$.
 Gauged Hamiltonian in terms of $ SL(2,R)$ for a  more general system has been discussed
in Ref. \cite{nr1}.
 
\subsection{Bender-Dunne (BD) Polynomials}

{\bf Case (i)}:   $ \alpha =l+1 =\beta  $.
We make a change of variable in Eq.(\ref{4}), 
$\cosh^{2}(x)=t$ , yielding
\bea\label{11}
&&t(t-1)\eta^{\prime\prime}(t) -[(l+3/2)-(2l+3-2ia)t-2iat^2]\eta^\prime (t)+ 
\nonumber \\ 
&&+(1/4)[\epsilon-iazt]\eta(t) =0 \,  
\eea 
On further substituting
\beq\label{12}
\eta(t)=\sum_{n=0}^{\infty} \frac{P_n(\epsilon) t^{n}}{n!\Gamma (n+l+3/2)}\,,
\eeq 
yields the three-term recursion relation satisfied by the polynomials,
 $ P_n(\epsilon)$
\bea\label{13}
P_{n+1}(\epsilon)&-&[\frac{\epsilon}{4}+n(n+2l+2-2ia)]P_n(\epsilon)
\nonumber \\ &+& ia(n+l+1/2)n[M-2l-2n-1]P_{n-1}(\epsilon)=0\,.
\eea 
 First few Polynomials are:
 \bea\label{14}
P_0(\epsilon)&=& 1\,, \nonumber \\
P_1(\epsilon)&=& \frac{\epsilon}{4} \nonumber \\
P_2(\epsilon)&=& \frac{\epsilon^2}{16}+\frac{\epsilon}{4}(2l+3-2ia) +ia(l+\frac
{3}{2})(2l+3-M)
\eea 

For the case,  $M=2l+3$, ~$ P_1(\epsilon)$ is the critical Polynomial\cite{bd}.
On demanding $P_1(\epsilon)=0 $ correctly yields the QES energy eigenvalue
as given by Eq. (\ref{7}). On the other hand, for  $M=2l+5$,~ 
 $ P_2(\epsilon)$ is the critical Polynomial, and demanding 
$P_2(\epsilon)=0 $ correctly yields the QES energy as given by Eq. (\ref{8}).

Following Bender and Dunne \cite{bd}, it is easy to compute the norm ($\gamma_n $) of the
n-th polynomial. We find
 \beq\label{15} 
\gamma_n = (4ai)^n n!\prod_{k=1}^n(k+l+1/2)(M-2k-2l-1)
\eeq 
Weight factors ($\omega_1 $ and $\omega_2 $ )\cite{bd}, for  $M=2l+5 $ are 

\bea\label{16}
\omega_1 &=& \frac{2l+3-2ai}{\sqrt{(2l+3)^2-4a^2}}+\frac{1}{2}\nonumber \\
\omega_2 &=& -\frac{2l+3-2ai}{\sqrt{(2l+3)^2-4a^2}}+\frac{1}{2}
\eea 
Moments of weight function is defined by
 \beq\label{17}
\mu_n = \int dE \omega(E)E^n \eeq
It is easily shown that the $n$-th moment, for large n, is proportional to 
$(M+a^2)^n $.

One can similarly study the properties of the Bender-Dunne polynomials in
the other three cases.

\section{ Periodic PT invariant QES system}

As has been shown in \cite{anti}, 
if  under  the anti-isospectral transformation 
$x \longrightarrow ix \equiv y $,  the potential
$v(x)\longrightarrow \bar{v}(y)$ and  if 
 the potential $v(x) $ has m QES levels with energy
 eigenvalue and eigenfunctions  $E_k \  (k=0,1,2 \cdots m-1)$ and  
 $\psi_k(x)$
 respectively then the energy eigenvalues of $\bar{v}(y)$ are given by 
 \beq \label{18}
\bar{E}_k =-E_{m-1-k}, \ \ \bar{\psi}_k(y) = \psi_{m-1-k} (ix) 
\eeq
Under this  anti-isospectral transformation,$x\longrightarrow ix \equiv 
\theta$, it is easily seen that the Hamiltonian (\ref{1}) goes over to
\beq\label{19}
H= p^2 + [a\cos (2\theta) -iM]^2 +\frac{l(l+1)}{\sin^2(\theta)}
+\frac{l(l+1)}{\cos^2(\theta)}\,,
\eeq 
with $-1<l<0$. In this case $U(x)= U(x+\pi)$. 
This complex Hamiltonian 
is invariant under combined Parity [ $x\longrightarrow \frac{\pi}{2}
-x$] and Time reversal [ $t\longrightarrow -t \ \& i\longrightarrow
-i $]. 
Explicit QES solutions can be obtained from the solutions of hyperbolic
case discussed in detail in section 2.1 by using the anti-isospectral 
transformation as given in  Eq. (\ref{18}). One can easily construct 
the BD polynomials 
and their properties for this case  by following the methods outlined in 
section 2.2.

\section{Real QES systems}

Before completing this paper, it may be worthwhile to point out that,  
starting from the Hermitian, QES, DSHG potential $V(x)=[a\cosh(2x)-M]^2$, one
can add several perturbing terms and the resulting Hamiltonian are all 
examples of QES systems. In this section, we consider three such perturbing
terms.  

{\bf Case I}: Perturbation term, $V_1=\frac{l(l+1)}{\sinh^2(x)}$ with 
$-1<l<0 $.

The combined perturbed system is described by the Hamiltonian,
\beq\label{20}
H_I= p^2 + [a\cosh (2x) -M]^2 +\frac{l(l+1)}{\sinh^2(x)}\, ,
\eeq
We show that for integral values of $M-l-1$ or $M+l$, this is a QES 
problem and one can 
obtain $p$ QES eigenstates in case $M=l+1+p+s$ or $M=-l+p+s$ respectively, 
where $p=1,2,3,...$ and $s=0,1$.

We substitute
\beq\label{21}
\psi(x) = e^{-\frac{a}{2}\cosh (2x)}\phi(x) \, ,
\eeq
in the Schro$\ddot{o}$dinger equation $H\psi= E\psi$ with $H$ as given by
Eq. (\ref{20}) and obtain
\bea\label{22}
&&\phi^{\prime\prime}(x) -2a\sinh (2x)\phi^\prime (x)+
[(E-M^2-a^2) \nonumber \\
&&+2(M-1)a\cosh (2x)-\frac{l(l+1)}{\sinh^2(x)} ]\phi(x) 
=0 \,  
\eea
On further substituting
\beq\label{23}
\phi = [\sinh(x)]^{\alpha} \eta\,,
\eeq
we obtain 
\bea\label{24}
&&\eta^{\prime\prime}(x) +2[\alpha \coth(x)-a\sinh (2x)]\eta^\prime (x)+ 
[E-M^2-a^2 \nonumber \\
&&+\alpha^2-2(M-1)a+4(M-\alpha-1)a\cosh^2 (x) ]\eta(x) =0 \,  
\eea
provided
\beq\label{25}
\alpha(\alpha-1)=l(l+1)\,.
\eeq
Eq.(\ref{25}) implies either  $\alpha=l+1$ or $\alpha=-l$.
We first consider $\alpha =l+1 $ and then it is easy to see that 
Eq. (\ref{24}) with $\alpha=l+1$ has p QES solutions in case $M=l+1+p+s$
where $s=0$ or $s=1$.
In particular the solutions are of the form
\beq\label{26}
\eta = [\cosh(x)]^{s}\sum_{n=0}^{\infty} a_n [\cosh^2(x)]^{n}\,,
\eeq
where $s=0$ in case $M=l+2p$ while $s=1$ in case $M=l+2p+1$. Few low
lying solutions are 
\bea\label{27}
&&M=l+2\,,~~\eta = constant\,,~~ E=a^2+2a(l+1)+2l+3\,, \nonumber \\
&&M=l+3\,,~~\eta=\cosh(x)\,,~~E=a^2+2al+2l+5\,,
\eea
\bea\label{28}
&&M=l+4\,,~~ E=y+a^2+3(2l+5)+2(l+3)a\,, \nonumber \\
&&\eta=A\cosh^2(x)+B\,,~~y=-2(l+2+2a)\pm 2\sqrt{(l+2+2a)^2-4a}\,.
\eea 
\bea\label{29}
&&M=l+5\,,~~ E=y+a^2+8(l+3)+2(l+4)a\,, \nonumber \\
&&\eta=A\cosh^3(x)+B\cosh(x)\,, \nonumber \\
&&y=-(4l+9+8a)\pm 2\sqrt{(l+2a)^2+3(2l+3)}\,.
\eea 

The results for $\alpha=-l$ are immediately obtained from above by replacing
everywhere $l$ by $-l-1$. 

By making the substitution $\cosh^2(x) =t$ in Eq. (\ref{24}), one can show that the corresponding
Bender-Dunne polynomials satisfy three term recursion relation.

{\bf Case II}: Perturbation term, $V_2=-\frac{l(l+1)}{\cosh^2(x)}$.

 The Hamiltonian of the system is thus given by
\beq\label{34}
H_2= p^2 + [a\cosh (2x) -M]^2 -\frac{l(l+1)}{\cosh^2(x)}\,,
\eeq
where $l$ is any real number. We again 
show that for integral values of $M-l-1$ or $M+l$, this is a QES problem and 
one can obtain $p$ QES eigenstates in case either $M=l+1+p+s$ or if 
$M=-l+p+s$ respectively, where $p=1,2,3,...$ and $s=0,1$.

We substitute
\beq\label{35}
\psi(x) = e^{-\frac{a}{2}\cosh (2x)} \cosh^\alpha (x) \eta, 
\eeq
in the Schro$\ddot{o}$dinger equation $H_2\psi= E\psi$ with $H_2$ as given by
Eq. (\ref{11}) to obtain
\bea\label{36}
&&\eta^{\prime\prime}(x) +2[\alpha \tanh(x)-a\sinh (2x)]\eta^\prime (x)+
[E-M^2-a^2 \nonumber \\
&&+2a(2\alpha-M+1)+\alpha^2+4(M-\alpha-2)a\cosh^2 (x) ]
\eta(x) =0 \,,  
\eea
provided
\beq\label{37}
\alpha(\alpha-1)=l(l+1)\,.
\eeq
This implies  $\alpha=l+1$ or $\alpha=-l$. 
let us consider first, $\alpha=l+1$. It is easy to see that 
Eq. (\ref{36}) with $\alpha=l+1$ has p qes solutions in case $M=l+1+p+s$
where $s=0$ or $s=1$. 
In particular the solutions are of the form
\beq\label{38}
\eta = [\sinh(x)]^{s}\sum_{n=0}^{\infty} a_n [\sinh^2(x)]^{n}\,,
\eeq
where $s=0$ in case $M=l+2p$ while $s=1$ in case $M=l+2p+1$. Few low
lying solutions are 
\bea\label{39}
&&M=l+2\,,~~\eta = constant\,,~~ E=a^2-2a(l+1)+2l+3\,, \nonumber \\
&&M=l+3\,,~~eta=\sinh(x)\,,~~E=a^2-2al+2l+5\,,
\eea
\bea\label{40}
&&M=l+4\,,~~ E=y+a^2+3(2l+5)-2a(l-1)\,, \nonumber \\
&&\eta=A\cosh^2(x)+B\,, \nonumber \\
&&y=-2(l+2+2a)\pm 2\sqrt{(l+2+2a)^2-4a(2l+3)}\,.
\eea 
\bea\label{41}
&&M=l+5\,,~~ E=y+a^2+8(l+3)-2a(l-2)\,, \nonumber \\
&&\eta=A\sinh^3(x)+B\sinh(x)\,, \nonumber \\
&&y=-(4l+9+4a)\pm 2\sqrt{(l-2a)^2+3(2l+3)}\,.
\eea 

It is easy to convince oneself that all the above solutions
are still solutions with replacement of $l$ by $-l-1$ everywhere. 

The three term recursion relations for the associate polynomials is obtained in this case by making the substitution $\sinh^2(x)=t$
in  Eq. (\ref{36}). 

{\bf Case III}: Perturbation term, 
$V_3=\frac{l(l+1)}{\sinh^2(x)}-\frac{g(g+1)}{\cosh^2(x)}$.

We now show that it is still a QES problem even if the perturbation is
the sum of the two perturbations considered in Case I and Case II, 
i.e. consider the Hamiltonian
\beq\label{46}
H_3= p^2 + [a\cosh (2x) -M]^2 +\frac{l(l+1)}{\sinh^2(x)}
-\frac{g(g+1)}{\cosh^2(x)}\,,
\eeq
where $-1 < l <0$ so that the singularity at $x=0$ is not strong enough. 
We show that for positive integral values of either $(M-l-g-1)/2$ or
$(M+l+g+1)/2$ or $(M+l-g)/2$ or $(M-l+g)/2$, this is a QES problem and one can 
obtain $p$ QES eigenstates in case either $M=l+g+2p+1$ or $M=-l-g+2p-1$ or
$M=-l+g+2p$ or $M=l-g+2p$, 
where $p=1,2,3,...$.  

We substitute
\beq\label{47a}
\psi(x) = e^{-\frac{a}{2}\cosh (2x)} [\cosh(x)]^{\alpha} 
[\sinh(x)]^{\beta} \eta \, ,
\eeq
in the Schro$\ddot{o}$dinger equation $H_3\psi= E\psi$ with $H_3$ as given by
Eq. (\ref{46}) 
we find that QES solutions exist only when $\alpha=g+1,-g$ and $\beta=l+1,-l$. 
On choosing $\alpha= g+1,\beta=l+1$, we obtain
\bea\label{47}
&&\eta^{\prime\prime}(x) +2[(l+1)\coth(x)+(g+1) \tanh(x)-a\sinh (2x)]
\eta^\prime (x) \nonumber \\
&&+ \left [y+z\cosh^2 (x) \right ] \eta(x) =0 \,,  
\eea
where
\beq\label{48}
y=E-M^2-a^2+(l+g+2)^2+2a(2g-M+3)\,,~~z=4a(M-l-g-3)\,.
\eeq
It is easy to see that 
Eq. (\ref{47}) has p QES solutions in case $M=l+g+2p+1$ with $p=1,2,3,...$.
In particular the solutions are of the form
\beq\label{49}
\eta = \sum_{n=0}^{\infty} P_n [\cosh^2(x)]^{n}\,,
\eeq
in case $M=l+g+2p+1$. Two of the low lying solutions are 
\beq\label{50}
M=l+g+3\,,~~\eta = constant\,,~~ E=a^2-2a(2g+3)+2l+2g+5\,,
\eeq
\bea\label{51}
&&M=l+g+5\,,~~ \eta=A\cosh^2(x)+B\,, \nonumber \\
&&y=-2(l+g+3+2a)\pm 2\sqrt{(l+g+3+2a)^2-4a(2g+3)}\,.
\eea 
The results for the remaining three cases are immediately obtained from here
by replacing $(l,g)$ with $(-l-1,-g-1)$, or with $(-l-1,g)$ or with
$(l,-g-1)$.

\section{Conclusion}

In this paper, we have shown that the known QES DSHG (and hence DSG) systems 
(both Hermitian and complex PT-invariant one) can be further
enlarged by adding perturbations and still it continues to be a QES system.
It will be interesting to look at other QES examples and obtain new QES
systems by adding suitable perturbating terms.

\end{document}